\documentclass[english]{article}
\usepackage[T1]{fontenc}
\usepackage[latin9]{inputenc}
\usepackage{geometry}
\usepackage{babel}
\usepackage{units}
\usepackage{amsmath}
\usepackage{amssymb}
\usepackage{cancel}
\usepackage{graphicx}

\makeatletter

\providecommand{\tabularnewline}{\\}

\date{}
\usepackage{breakurl, color, babel, textcomp,graphicx, setspace, lineno, lmodern, hyperref}
\hypersetup{
colorlinks=true,
linkcolor=red,
citecolor = blue,
urlcolor=blue,
pdfpagemode=FullScreen}
\@ifundefined{showcaptionsetup}{}{%
 \PassOptionsToPackage{caption=false}{subfig}}
\usepackage{subfig}
\makeatother

\begin{document}
\title{Constriction Percolation Model for Coupled Diffusion-Reaction Corrosion
of Zirconium in PWR}
\author{Asghar Aryanfar$^{*,\ddagger}$, William Goddard III$^{*}$, Jaime
Marian$^{\dagger}$}
\maketitle
\begin{center}
\emph{$*$California Institute of Technology, 1200 E California Blvd,
Pasadena, CA 91125 }
\par\end{center}

\begin{center}
\emph{$\dagger$University of California, 410 Westwood Plaza, Los
Angeles, CA 90095 }
\par\end{center}

\begin{center}
\emph{$\ddagger$Bahçe\c{s}ehir University, 4 Ç\i ra\u{g}an Cad, Be\c{s}ikta\c{s},
Istanbul, Turkey 34349}
\par\end{center}
\begin{abstract}
Percolation phenomena are pervasive in nature, ranging from capillary
flow, crack propagation, ionic transport, fluid permeation, etc. Modeling
percolation in highly-branched media requires the use of numerical
solutions, as problems can quickly become intractable due to the number
of pathways available. This becomes even more challenging in dynamic
scenarios where the generation of pathways can quickly become a combinatorial
problem. In this work, we develop a new constriction percolation paradigm,
using cellular automata to predict the transport of oxygen through
a stochastically cracked Zr oxide layer within a coupled diffusion-reaction
framework. We simulate such branching trees by generating a series
porosity-controlled media. Additionally, we develop an analytical
criterion based on compressive yielding for bridging the transition
state in corrosion regime, where the percolation threshold has been
achieved. Our model extends Dijkstra\textquoteright s shortest path
method to constriction pathways and predicts the arrival rate of oxygen
ions at the oxide interface. This is a critical parameter to predict
oxide growth in the so-called post-transition regime, when bulk diffusion
is no longer the rate-limiting phenomenon.

\end{abstract}

\section{Introduction}

The corrosion and fracture of zirconium clad in the presence of high-temperature
water is the main failure mechanism in cooling pipelines of pressurized
water reactors (PWR). \cite{Motta_15,Cox_05,Cox_90,Motta_07} The
gradual oxidation is the result of diffusion of oxygen into the depth
of metal matrix, followed by chemical reaction in the corrosion front.
Several studies have shown that the oxide scale grows as cubic law
versus time during pre-transition period ( $\sim t^{\nicefrac{1}{3}}$)
as opposed to typical parabolic diffusion behavior ($\sim t^{\nicefrac{1}{2}}$).
\cite{Sabol_75,Forsberg_95,Cox_60} The oxygen diffusion into metallic
structure leads to large augmentation in volume and internal compressive
stresses due to Pilling-Bedworth ratio.\footnote{$R_{PB}=\cfrac{V_{ox}}{V_{Zr}}=\cfrac{\rho_{ox}}{\rho_{Zr}}\approx1.56$
where $V$ and $\rho$are the molar volume and $\rho$ is mass density
respectively.} \cite{Lustman_55} The fracture reason is attributed to the residual
stresses from cyclic cooling, embrittlement from hydrides precipitation
and phase transformation during non-stoichiometric oxidation of zirconium
as well as yielding due to compressive and the balancing tensile stresses.\cite{Puchala_13,Platt_14,Cox_05}.\footnote{From tetragonal to monoclinic and to the cubic phase.}
The randomly-distributed cracks are merely sensitive to original spatial
distribution/concentration of defects/grain boundaries \cite{Chen_06,Adamson_07}.
Consequently water can penetrate into the cracks and the oxygen gets
easy access to corrosion sites without the original pre-cracking diffusion
barrier. This event leads to jump in corrosion kinetics \cite{Cox_60}.
The diffusion process via grain boundaries and material matrix (i.e
lattice) has been studied in the context of percolation \cite{Grimmett_13}.
One of the illustrative methods for percolation is cellular automata
paradigm which is typically studied in the two distinct context of
site and bond percolation. \cite{Chopard_89,Chopard_91,Newman_01}
Later studies have complemented the percolation with reaction in the
diffusion front. \cite{Chopard_94} However, the precise quantification
of diffusion through the shortest constriction pathways, particularly
during the oxidation process and wide range of percolation regime,
distinguished by fracture has not been addressed before. In this paper,
we develop a coupled diffusion-reaction framework, based on the two
percolation paradigms for predicting the corrosion rates after initiation
of cracks. The constriction and tortuous geometry of percolation pathways
as well as the reactive term has central role in our model for predicting
the ultimate corrosion kinetics. 

\section{Methodology}

\subsection{Approach}

The inhomogeneous percolation of the oxygen within the oxide scale
could either be interpreted as diffusion within the cracked network
during post-transition regime or within the grain boundaries or material
imperfections during the pre-transition development. In fact such
two patterns are highly correlated as the cracking is the most feasible
to occur through relatively weaker grain boundaries/imperfections.
During the initial stage of oxidation, the abundant oxygen from electrolyzed
water reacts with the zirconium metal and therefore the corrosion
is in fact \emph{reaction-limited}. However, after a sufficient penetration
of oxide scale into the depth, the reaction front suffers from \textquotedblleft breathing\textquotedblright{}
and corrosion kinetics turns to be dominantly \emph{diffusion-limited}.
Upon the fracture, the cracks propagate in columnar shape, preferably
along the weakest shear bands (i.e. grain boundaries) and the water
gets easy access to the oxidation front (i.e. oxide/metal interface). 
\begin{center}
\begin{figure}
\begin{centering}
\includegraphics[width=1\textwidth]{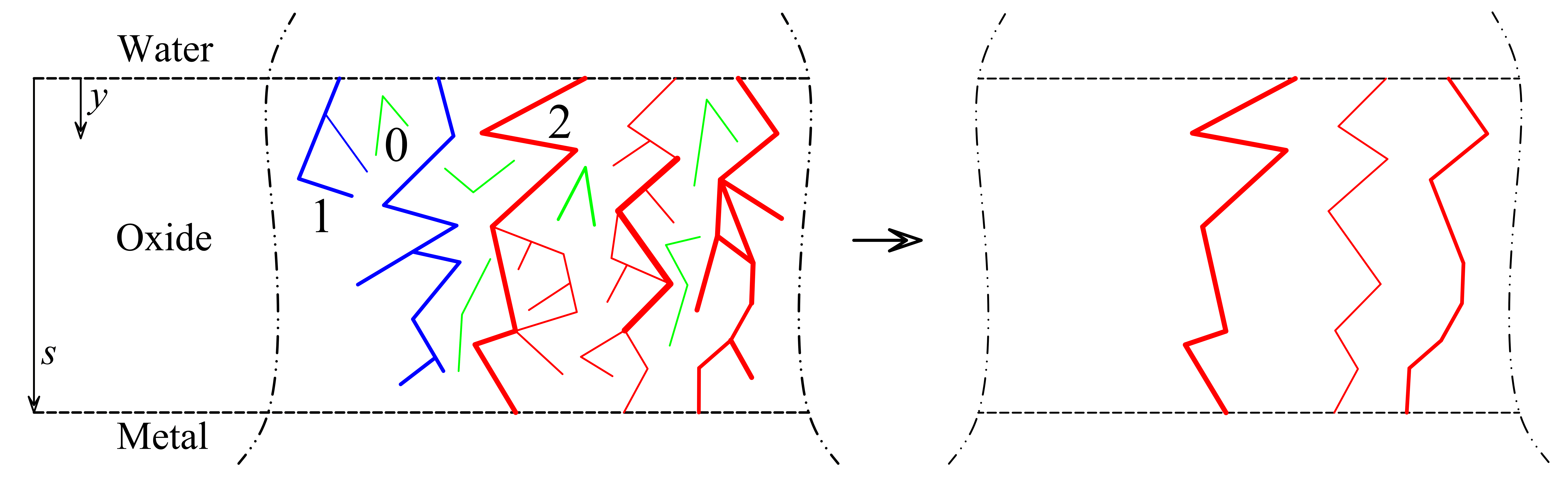}
\par\end{centering}
\caption{Schematics of cracks network (left) and their corresponding constriction
rivers (right). \label{fig:Cracks}}
\end{figure}
\par\end{center}

\subsection{Percolation clusters}

\subsubsection{Characterization}

Given a general network of cracks in Figure \ref{fig:Cracks}, for
the purpose of simulation we can differentiate each one either via
the connection to the either of the interfaces, the constriction value
with the inherent tortuosity as below: 

\textbf{0. Islands:} These confined areas have no access to the any
of the interfaces. Therefore their role can be neglected.

\textbf{1. Partial Cracks:} These cracks need have partial progress
within the oxide layer from the water/oxide interface. The transport
of water occurs through the tortuous crack, while the rest of the
diffusion within the oxide occurs through the shortest path. (i.e.
straight line.) 

\textbf{2. Full cracks: }These cracks provide full connection between
the water/oxide and oxide/metal interfaces. 

The tortuous geometry of each crack not only elongates the transport
route, but also provides a projection for the transport flux. Therefore
the diffusion coefficient for each crack $D_{cr}$is expressed as: 

\begin{equation}
\frac{\tau_{cr}^{2}}{D_{cr}}=\frac{\tau_{ox}^{2}}{D_{ox}}+\frac{\tau_{w}^{2}}{D_{w}}\label{eq:CrackSeries}
\end{equation}

where $D$ and $\tau$are the  diffusion coefficient and tortuosity
of crack, oxide and water respectively. Comparison the diffusivity
values for oxide scale $D_{ox}\approx10^{-17}m^{2}/s$ \cite{Youssef_14}
and water $D_{w}\approx10^{-8}m^{2}/s$ \cite{Holz_00} leads to: 

\begin{equation}
D_{ox}<D_{part}\ll D_{full}<D_{w}\label{eq:DComparison}
\end{equation}

where $D_{part}$ and $D_{full}$ are the diffusion coefficient values
for the partial and full cracks. 

The significant difference in Equation \ref{eq:DComparison} addresses
that there is a jump in the oxide growth rate when the porosity of
crack network reaches the percolation threshold value $p=p_{c}$.
Consequently, the homogenized diffusivity $D_{EFF}$ during post-transition
period can be simplified in $2D$ as: 

\begin{equation}
D_{EFF}\approx\frac{D_{w}}{L}\sum_{k=1}^{n}\frac{l_{k}}{\tau_{k}^{2}}\label{eq:DDiscretized}
\end{equation}

\begin{center}
\begin{figure}
\begin{centering}
\subfloat[PseudoFlowchart for cracks.\label{fig:FlowChart}]{\begin{centering}
\includegraphics[height=0.17\textheight]{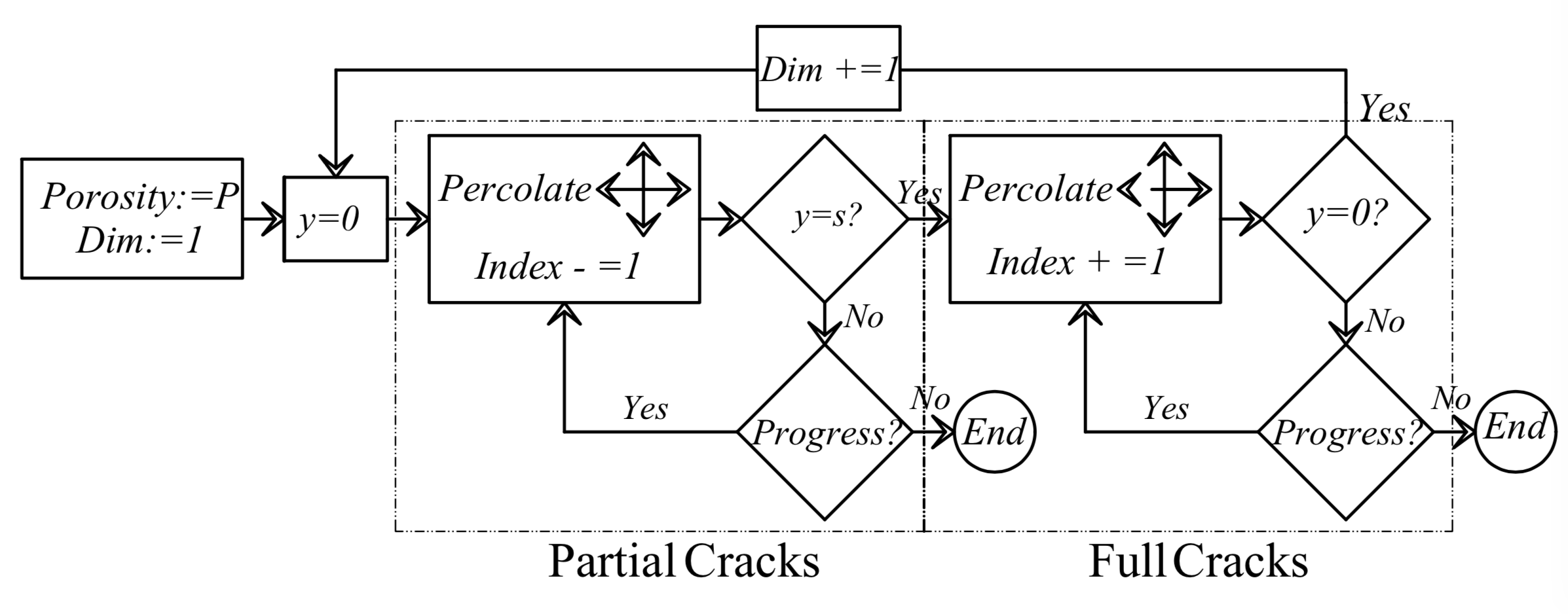}
\par\end{centering}
}\hfill{}\subfloat[Crack density.\label{fig:M}]{\centering{}\includegraphics[height=0.16\textheight]{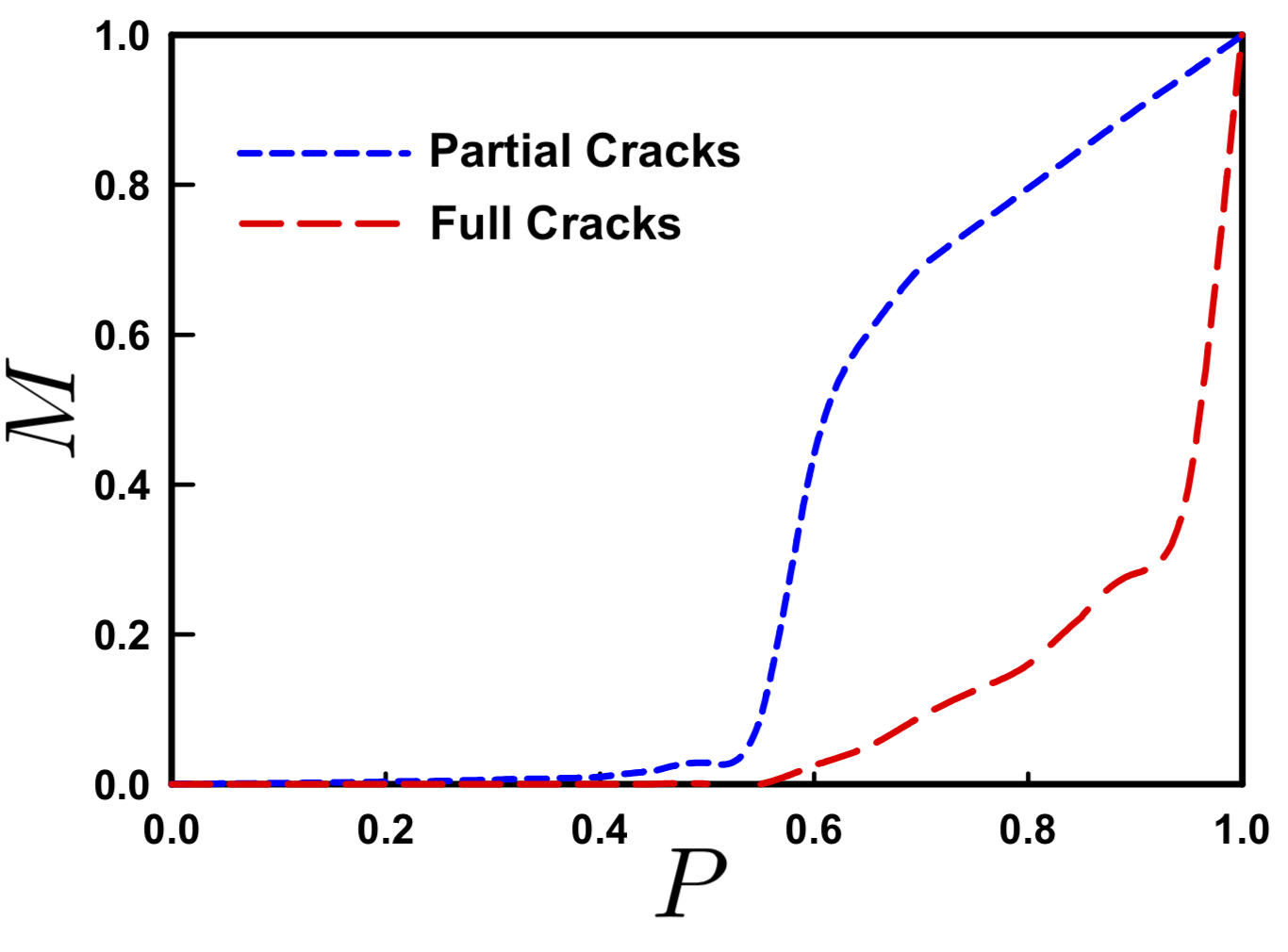}}
\par\end{centering}
\caption{Computational development chart (left) and crack density (right). }
\end{figure}
\par\end{center}

We simulate such medium by generating stochastic binary medium with
the developing porosity in time. Utilizing the cellular automaton
paradigm, we extend Dijkstra\textquoteright s shortest path algorithm
\cite{Dijkstra_90}, for extracting the \emph{Constriction rivers}
(\emph{CRs}). The diffusion through cracks has been simulated by the
square bond percolation ($4^{2}$), accompanied with the site percolation
method for comparison and verification. The percolation threshold
probability $p_{c}$, which divides the pre and post-transition growth
regimes, for the former is known to be $\approx0.5$ while in the
latter is $\approx0.5928$, after which some of the partially-formed
cracks tend full cracks. \cite{Stauffer_94,Deng_05,Jacobsen_14}

Figure \ref{fig:FlowChart} explains the computational algorithm for
extracting the constriction pathways illustrated in Figure \ref{fig:Cracks}.
We summarize the procedure as below: 

\textbf{i. Forward percolation:} Starting from the water/oxide interface
($y=0$), percolate forward from the $1^{st}$ order neighbors and
index each new addition in descending order. Such index tangibly correlates
with the amount of time the water has reached that location. For full
cracks, the percolation will reach to the oxide/metal interface ($y=s$). 

\textbf{ii. Backward Percolation:} The largest index in the oxide/metal
interface ($y=s$) indicates that water has reached there the earliest.
Therefore, starting from that element, we revert backwards from the
$1^{st}$ order neighbors by ascending order of indexes until reaching
back the water/metal interface ($y=0$). The extracted path is the
shortest distance between two interfaces. The usability of pores within
the cracked medium depends on if they have been captured as a part
of the shortest path. For the pathways of the same beginning/end (i.e.
same length) one of them is eliminated in favor of the other. 

\textbf{iii. Constriction Percolation: }The constriction along each
river would control the diffusion and the flux of water. In order
to capture that, we periodically increase the thickness ($Dim$) in
2D (i.e. cube in 3D) in cellular automata paradigm. Additionally starting
from shortest river, obtained in the first two steps ensures the shortest
path for the thickest possible river as well.

Figure \ref{fig:M} illustrates the density of states ($M$) for the
obtained partial and full cracks based on original porosity and Figure
\ref{fig:Sample} is a sample illustration based on square site percolation
($4^{2}$) beyond percolation limit. ($p>p_{c}$). The black routes
are the only top-to-bottom connection pathways and the color map value
on each site correlates with the time oxygen has reached that location.
We will elaborate on this further in 
\begin{center}
\begin{figure}
\begin{centering}
\subfloat[A site percolation sample. $p=0.55$.\label{fig:Sample}]{\begin{centering}
\includegraphics[height=0.22\textheight]{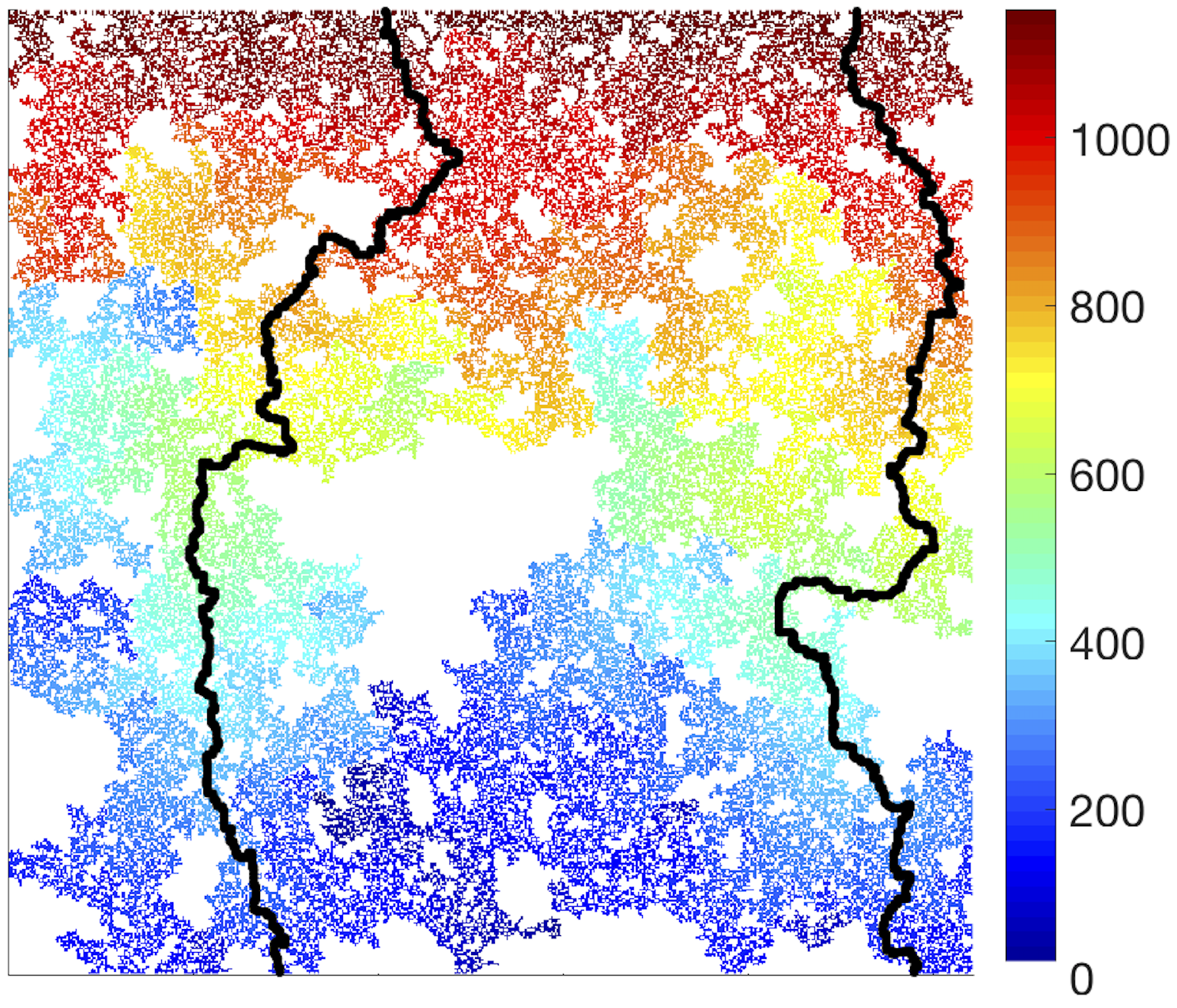}
\par\end{centering}
}\hfill{}\subfloat[Original image.\label{fig:Original}]{\centering{}\includegraphics[height=0.22\textheight]{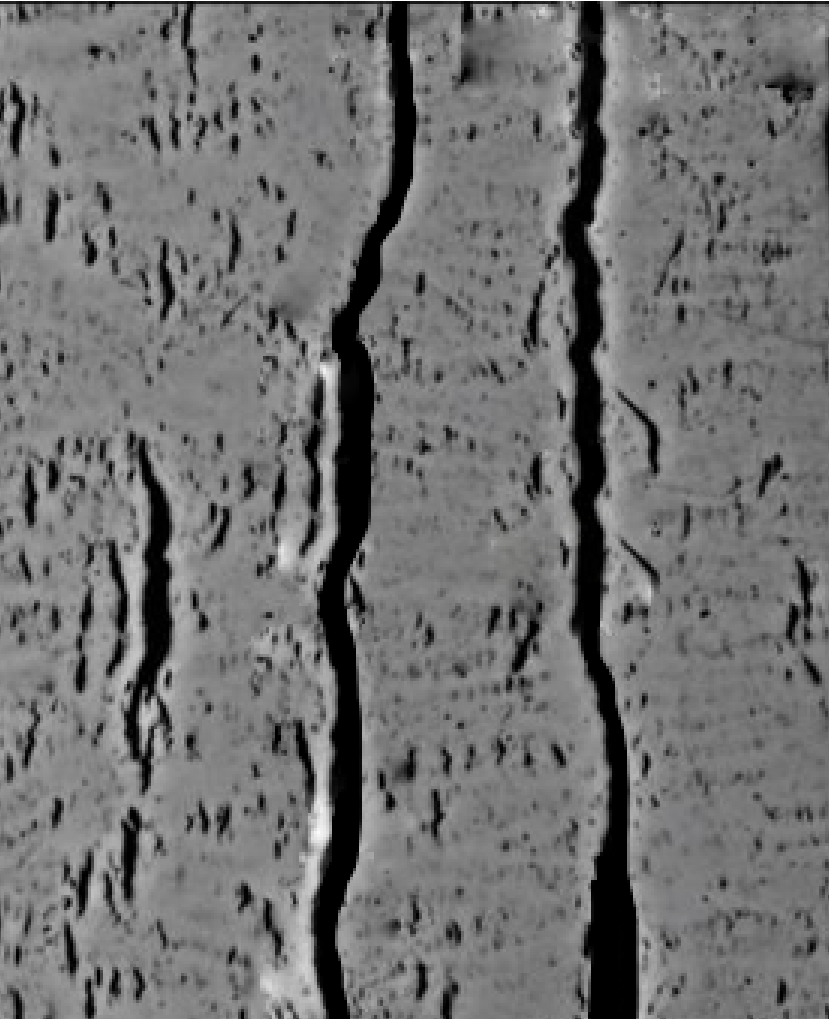}}\hfill{}\subfloat[Illustration of $CRs$ (in red).\label{fig:Effective}]{\centering{}\includegraphics[bb=0bp 0bp 2100bp 2454bp,height=0.22\textheight]{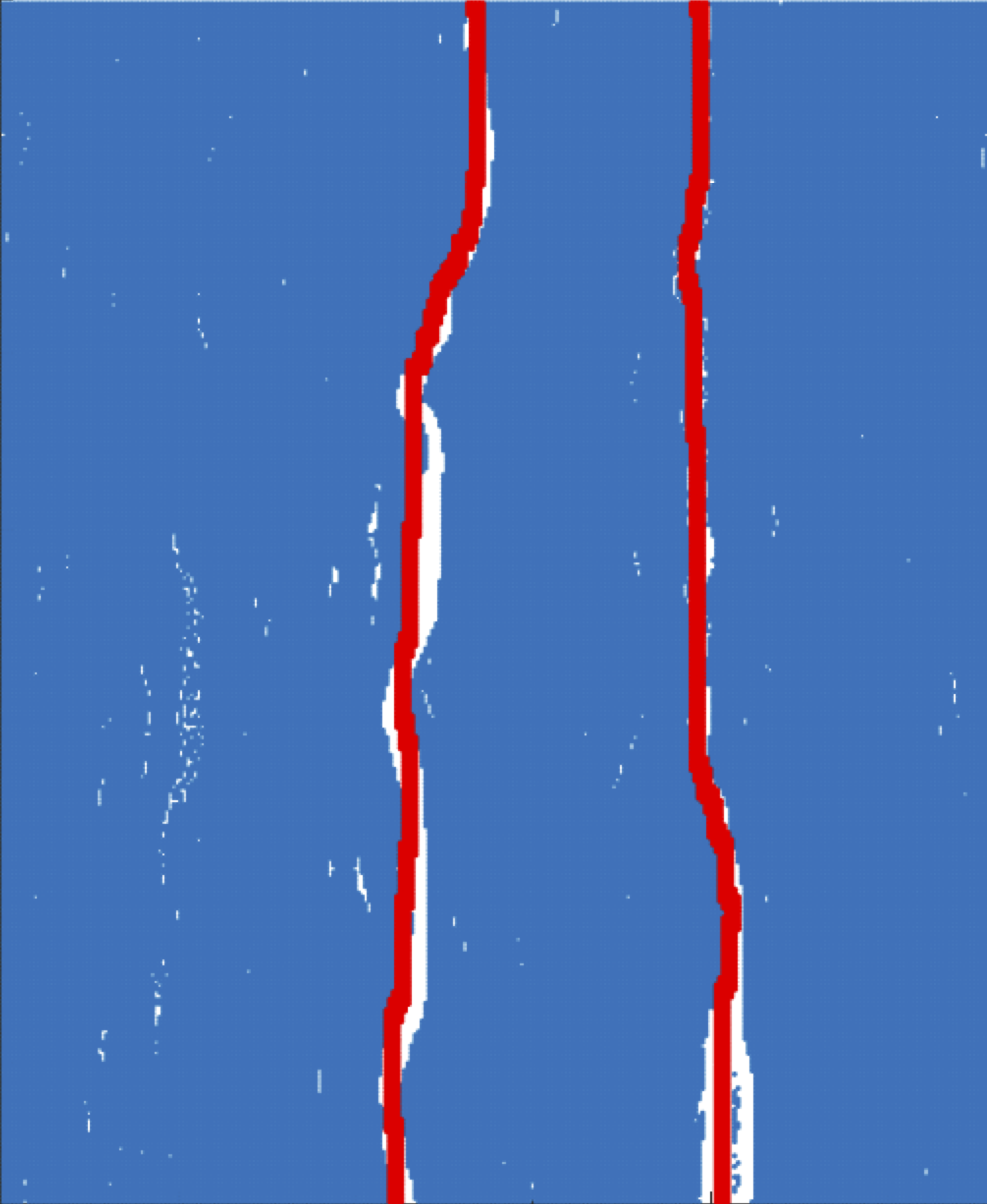}}
\par\end{centering}
\caption{\textbf{(a)} Sample site percolation: blue: cracked network, green:
partial cracks, red: full cracks. \textbf{(b) }Original cracked oxide
image. \textbf{(c) }Extracted constriction rivers (blue: metal, white:
voids, red: \emph{CRs}). }
\end{figure}
\par\end{center}

The extraction of \emph{CRs }from the given cracked medium would be
possible by implementing binarizaion on the original grayscale image
(Figure \ref{fig:Original}) via Otsu's method \cite{Otsu_75}. This
could be possible choosing a threshold such to minimize the intra-class
variance $\sigma^{2}$ defined as below: 
\begin{center}
minimize $\sigma^{2}$ such that: 
\par\end{center}

\vspace*{-1.5cm}

\begin{center}
\begin{equation}
\text{}\begin{cases}
\sigma^{2}=\omega_{1}\sigma_{0}^{2}+\omega_{2}\sigma_{1}^{2}\\
\omega_{1}+\omega_{2}=1
\end{cases}\label{eq:Otsu}
\end{equation}
\par\end{center}

where $\sigma_{0}^{2}$ and $\sigma_{1}^{2}$ are the variance for
the divided black and white groups and $\omega_{1}$ and $\omega_{2}$
are their corresponding fraction. Performing the procedure in flowchart
\ref{fig:FlowChart} on the original image (Figure \ref{fig:Original})\footnote{Taken from experimental image at PNNL.},
the $CRs$ could be obtained as shown in Figure \ref{fig:Effective}
The real-time simulation of extracting of constriction percolation
pathways is shown in the supplemental materials.\footnote{Also available here: \href{https://www.youtube.com/watch\%3Fv\%3D82lAUKEcqS0}{https://www.youtube.com/watch?v=82lAUKEcqS0} }

Furthermore the tortuosity of cracked pathways can be calculated from
the extracted $CRs$, the average tortuosity $\bar{\tau}$ for site
and bond percolations are shown versus original porosity in 2 and
3 dimensions in Figure \ref{fig:TortPor}. The computed diffusion
coefficients from Equation \ref{eq:DDiscretized} is shown in Figure
\ref{fig:DiffPor} respectively for partial ($p<p_{c}$ ) and full
($p>p_{c}$) cracks. 
\begin{center}
\begin{figure}
\subfloat[Tortuosity.\label{fig:TortPor}]{\centering{}\includegraphics[height=0.24\textheight]{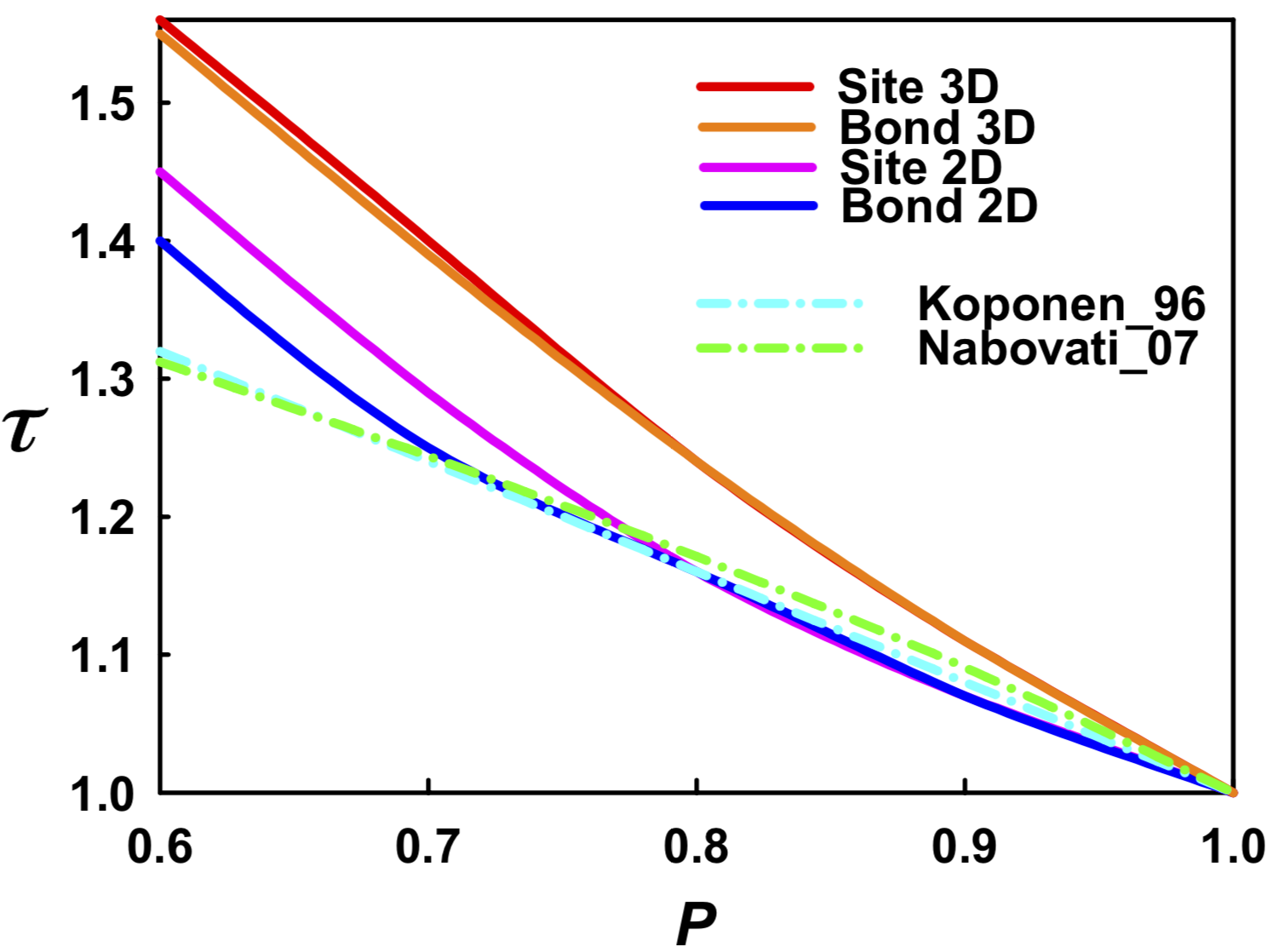}}\hfill{}\subfloat[Diffusion coefficient.\label{fig:DiffPor}]{\centering{}\includegraphics[height=0.25\textheight]{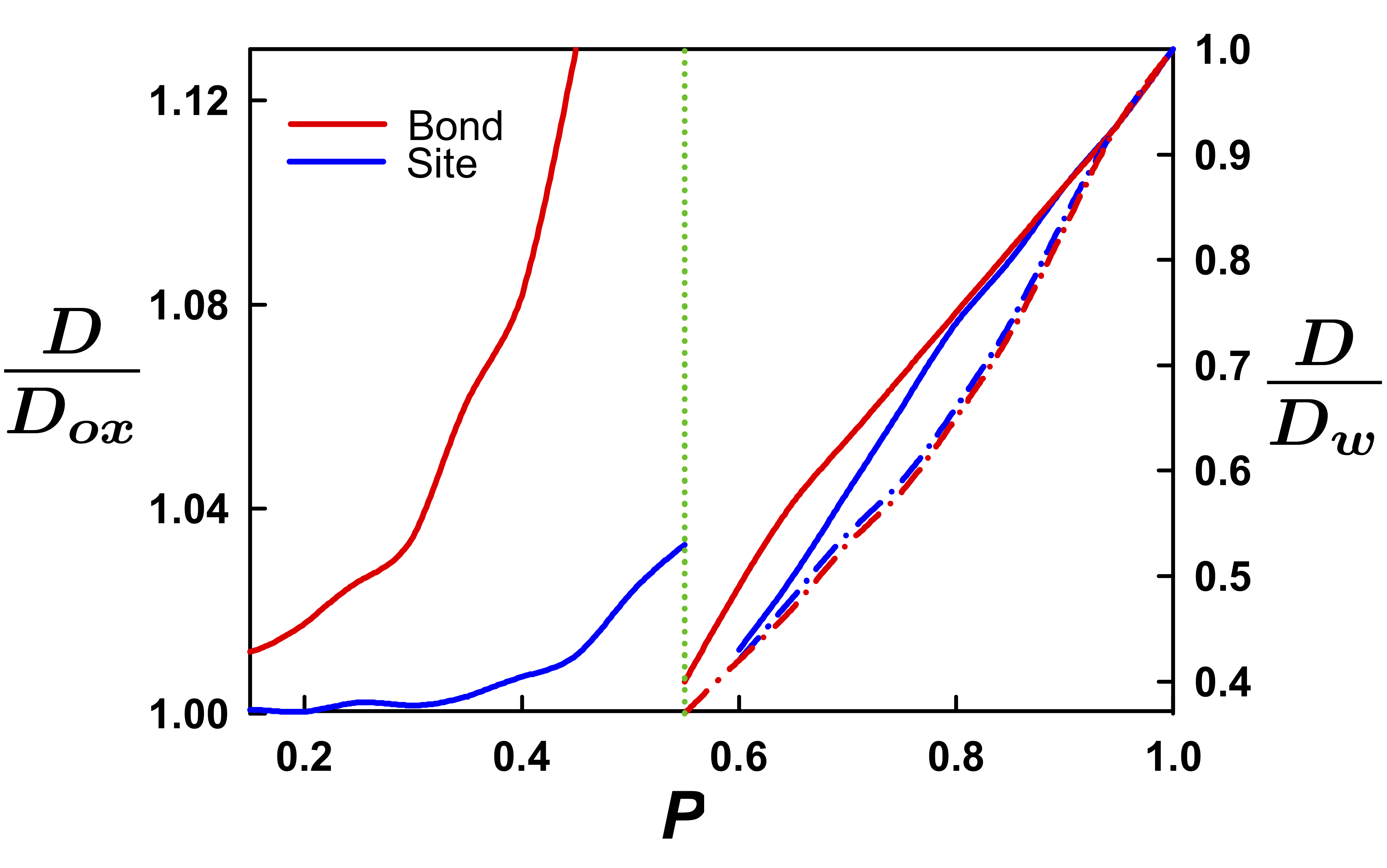}}

\caption{The geometric role of cracks on the diffusivity. }
\end{figure}
\par\end{center}

\subsubsection{Scaling Dimension }

Scaling dimension $\alpha$, in fact represent the scaling role of
the percolating cluster versus the dimension of the medium. In other
words, for the percolation paradigm with the domain scale $L$, there
is a power coefficient $\alpha\in{\rm R}$ for which the density of
states for percolating cluster $M(L)$ correlates with the domain
scale as: 

\begin{equation}
M(L)\propto L^{\alpha}\label{eq:DOS}
\end{equation}

We have performed the constriction percolation paradigm in square
bond paradigm ($4^{2}$), given in flowchart \ref{fig:FlowChart}
from the center of the medium, for various scales. Upon reaching the
threshold (i.e. two facing boundaries) the computations has been stopped.
The density of states for the connecting percolation pathways $M(L)$
has been plotted against the domains scale $L$in Figure \ref{fig:Scaling},
versus the exponent limits given in the literature. \cite{Stauffer_94}.
To ensure the statistical converges, each simulation point is the
average of 10 stochastic computations. Additionally, Figure \ref{fig:ScalingSample}
visualizes a sample percolation computation for the domain scale of
500. \footnote{Note that the the parameters are dimensionless. ($[]$), }
\begin{center}
\begin{figure}
\subfloat[Scale dependency of density of states (M) based on site and bond percolation.
\label{fig:Scaling} ]{\centering{}\includegraphics[height=0.3\textheight]{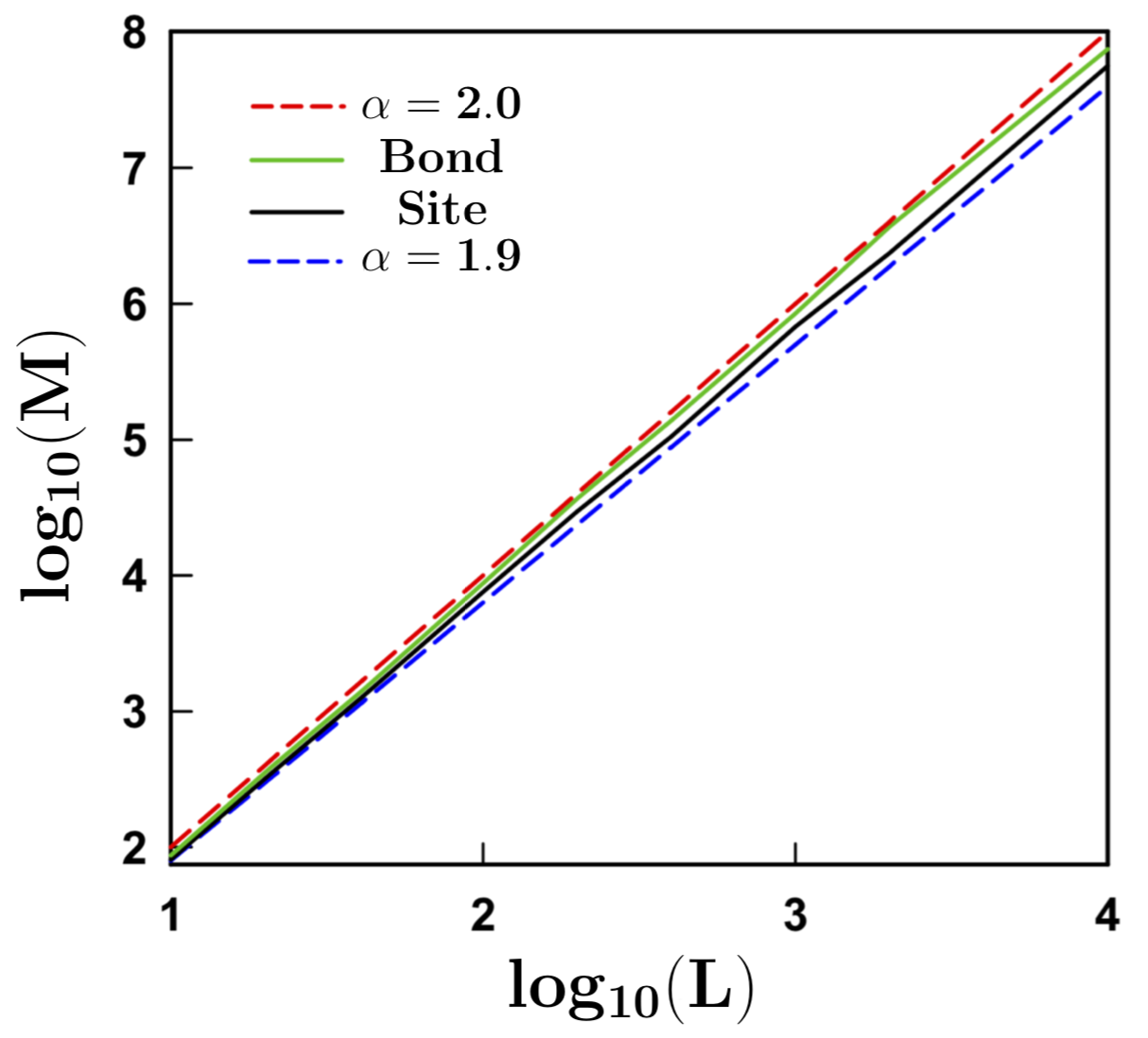}}\hfill{}\subfloat[A bond percolation from center. ($p=0.65$)\label{fig:ScalingSample}]{\centering{}\includegraphics[height=0.32\textheight]{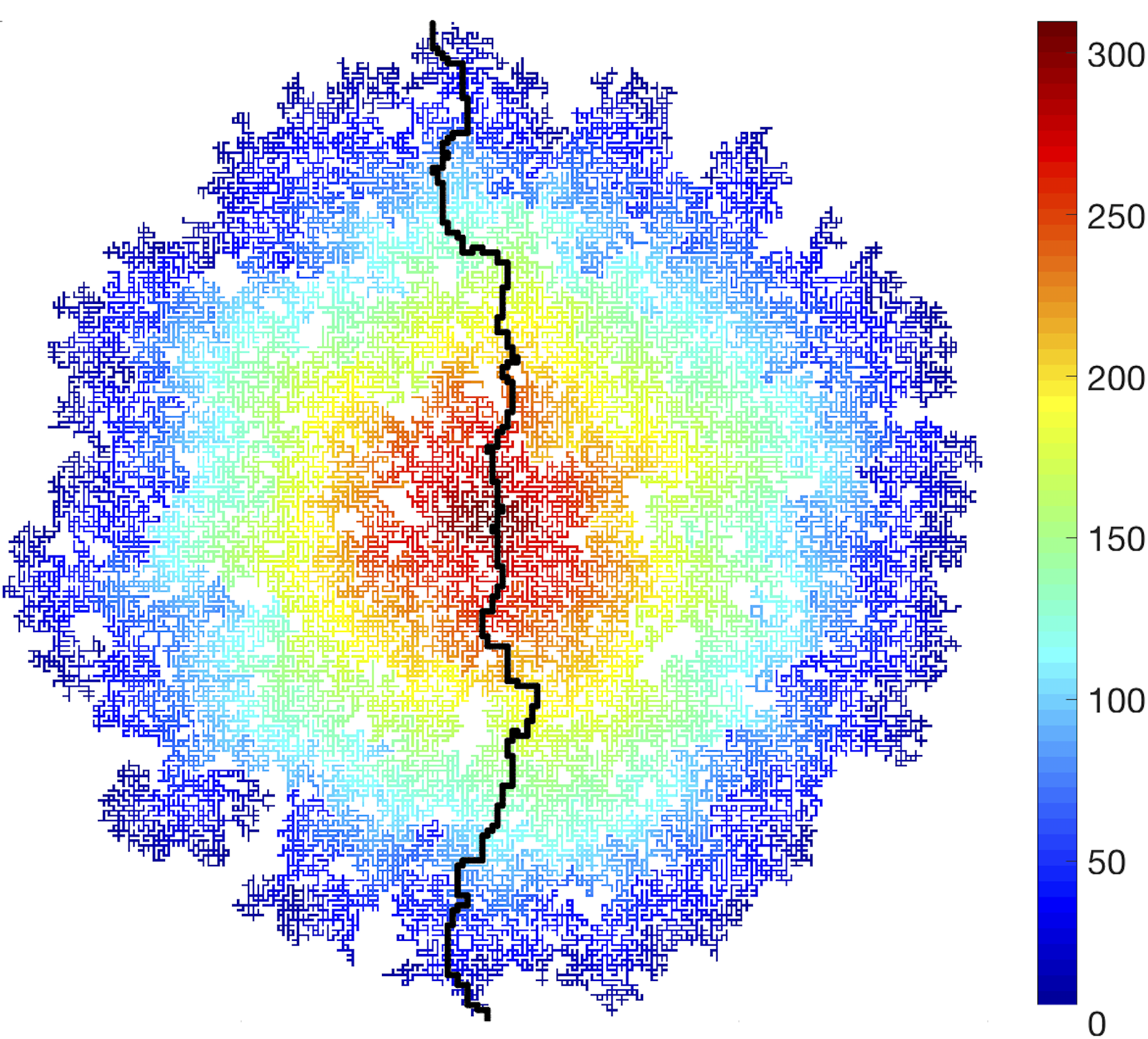}}

\caption{Scale studies. }
\end{figure}
\par\end{center}

\subsection{Formulation}

During the oxidation process, initially the oxygen from the water
electrolysis starts filling in the zirconium matrix until reaching
the stoichiometric limit, where the zirconium dioxide forms. Subsequently
the oxide/metal (i.e. reaction site) growth deeper within the metal
during so-called pre-transition regime. However, after growing to
a sufficient extent, the fracture gradually occurs and the cracks
accumulate and propagate up to the corrosion front. Therefore, the
transport of oxygen dominantly occurs via water percolation within
the crack network, leading to a sudden jump in corrosion kinetics.

Merging two growth regimes, the evolving oxygen concentration ($O$)
is given via the extended diffusion Equation in $1D$ as\cite{ARYANFAR_16}:\footnote{For simplicity, the diffusion due to pressure is neglected.}

\begin{equation}
\frac{\partial O}{\partial t}=D(T)\left(\frac{\partial^{2}O}{\partial y^{2}}+\Delta(T)\frac{\partial O}{\partial y}\right)-kO\label{eq:Diff}
\end{equation}

where $y$ is the depth variable and $t$ is the time defined in Figure
\ref{fig:Cracks}. The extra term in the $RHS$ of Equation \ref{eq:Diff}
represents the thermomigration. According to the Suret-ludwig effect,
the vacancy mediated motion of substitutional atoms is expected to
occur down the temperature gradient, while the interstitial solute
move in the opposite direction. \cite{Rahman_14} The second extra
term represents the consumption of oxygen due to oxidation process
and $k$ is the reaction constant. 

The diffusivity $D(T)$ is regime-dependent and defined as below: 

\begin{equation}
D(T)=\begin{cases}
{\displaystyle \Large D_{0,ox}\exp\left(\cfrac{-Q_{ox}}{RT}\right)} & \text{Pre}\\
{\displaystyle \Large D_{0.w}\left[\frac{T}{T_{s}}-1\right]^{\gamma}} & \text{Post}
\end{cases}\label{eq:D}
\end{equation}

where the first Equation shows the Arrhenius-type relationship \cite{Youssef_14}
and the second Equation represents Speedy-Angell power law self-diffusion
of water.\cite{Holz_00}

Therefore the $\Delta(T)$ is obtained from chain derivation as: 

\begin{equation}
\Delta(T)=\begin{cases}
{\displaystyle \Large\frac{Q_{ox}}{RT^{2}}\frac{dT}{dy}} & \text{Pre}\\
{\displaystyle \Large\frac{\gamma}{T-T_{s}}\frac{dT}{dy}} & \text{Post}
\end{cases}\label{eq:alpha}
\end{equation}

There is presumably no oxygen in the zirconium matrix at the beginning,
therefore the initial condition would be: 

\begin{equation}
O(y,0)=0\label{eq:IC}
\end{equation}

On the other hand, on the water/oxide interface there is constant
concentration of oxygen provided from water radiolysis ( $O_{0}$)
: 

\begin{equation}
O(y,0)=O_{0}\label{eq:BC1}
\end{equation}

$O_{0}$ has been considered as the molar value of oxygen in the water.
(Table \ref{tab:Parameters}).\footnote{Oxygen from water: $O_{0}=\cfrac{1}{16}\cfrac{mol}{g}=\cfrac{1}{16}\cfrac{mol}{g}\times\cfrac{1g}{cm^{3}}\times\cfrac{1000cm^{3}}{L}=62.5M$ }

and no oxygen can escape from clad into the fuel side: (i.e. $J_{O}(L,t)=0$)
\cite{Reyes_18}

\begin{equation}
\frac{\partial O}{\partial y}(L,t)=0\label{eq:BC2}
\end{equation}

Generally during the diffusion, the oxygen should migrates inside
the zirconium matrix, however as the reaction occurs much faster rate
than the diffusion ($k\gg D$), the entire diffused oxygen reacts
towards the formation of oxide scale upon reaching the reaction sites
(i.e. diffusion front). Therefore the effective depth of the oxide
layer at a given time could be obtained by leveling the entire diffused
oxygen it with the stoichiometric saturation value of zirconium metal
to oxide as: 

\begin{equation}
s(t)=\frac{1}{2Z_{0}}\int_{0}^{t}O(y,t)dy\label{eq:s}
\end{equation}

The coefficient of 2 is due to stoichiometric ratio of oxygen to zirconium
(i.e. $\nicefrac{N_{O}}{N_{Zr}}=2$).\footnote{Total zirconium in action: $Z_{0}=\cfrac{1}{91}\cfrac{mol}{g}=\cfrac{1}{91}\cfrac{mol}{g}\times\cfrac{6.52g}{cm^{3}}\times\cfrac{1000cm^{3}}{L}=71M$ } 

\subsubsection{Transition state: }

The fracture in the zirconium oxide, is attributed to few factors,
such as the residual stresses during heating/cooling cycles and the
compressive stresses mainly via shear bands \cite{Cox_60,Allen_12}.
Here, we develop a simple formulation based on the latter to describe
the onset of cracking throughout oxidation development. The treat
the oxide scale as an epitaxial layer on the substrate metal surface,
exposed to biaxial stresses. Assuming the original cross sectional
area of zirconium metal $A_{0}$ and oxide depth $L$ , the corresponding
oxide volume would be $V=AL$ . Adding the increment of $ds$ to the
oxide depth would lead to increment of volume as $dV=Ady+ydA$ . Due
to confined boundaries from lateral dimensions we have $ydA\approx0$,
and therefore the growth can be approximated as 1D. 

Having one dimensional growth, the addition of the infinitesimal depth
$ds$ into the oxide layer would increase the total depth by $R_{PB}dy$
. ($ds=R_{PB}dy)$ and reduce the zirconium depth by $dy$. The variation
in total thickness $(dy)_{tot}$ consequently is: 

\[
(dy)_{tot}=(dy)_{ox}-(dy)_{zr}=ds-\cfrac{ds}{R_{PB}}=(\cfrac{R_{PB}-1}{R_{PB}})ds
\]

The relative volume variation becomes: 

\[
\cfrac{dV}{V}=\cfrac{dy}{y}=\cfrac{(R_{PB}-1)ds}{R_{PB}L+(R_{PB}-1)s(t)}
\]

where ($R_{PB}-1$) could also be tangibly interpreted as the \emph{addition
coefficient} in $1D$. Subsequently, for the biaxial loading, the
compressive stress is twice the homogeneous pressure $\sigma=2P$.
Thus the homogeneous stress within the oxide scale could be predicted
from real-time computation as:

\begin{equation}
\sigma(t)=2K(R_{PB}-1)\int_{0}^{s(t)}\frac{d\omega}{R_{PB}L+(R_{PB}-1)\omega(t)}\label{eq:Syc}
\end{equation}

where $K$ is the bulk modulus. On the verge of fracture, the stress
reaches the compressive yield limit in the oxide medium.\footnote{We treat the oxide-metal scale as a whole composite medium where the
fracture occurs within oxide compartment. } Upon reaching the transition moment we have: $\sigma(t)=S_{yc}$
and $s=s_{c}$. Therefore, solving the Equation \ref{eq:Syc} analytically
leads to: 

\begin{equation}
s_{c}=\frac{R_{PB}L}{R_{PB}-1}\left(\exp\left(\frac{S_{yc}}{2K}\right)-1\right)\label{eq:sc}
\end{equation}

The range values from table \ref{tab:Parameters} for compressive
yield stress and the bulk modulus for zirconium oxide are $S_{yc}\in[1.2-5.2]GPa$
and $K\in[72-212]GPa$, therefore the transition range of oxide scale
would be: 

\begin{equation}
s_{c}\in[0.4-5.1]\mu m\label{eq:scValue}
\end{equation}

The simulation parameters for oxide scale growth are shown in Table
\ref{tab:Parameters}.\footnote{The values of compressive yield strength $S_{yc}$ and the bulk modulus
$K$ are obtained from: \href{https://www.azom.com/properties.aspx\%3FArticleID\%3D133}{https://www.azom.com/properties.aspx?ArticleID=133} }$^{,}$\footnote{The value of reaction constant is considered to be corresponding to
the average grain size of $50nm$ in Ref. \cite{Zhang_07}, Figure
2 and considering the sample height: $l=0.1m$ as below: $k=4\times10^{-9}mg^{2}.dm^{-2}.d^{-1}=4\times10^{-9}\times10^{-2}\nicefrac{g}{m^{2}}\times(24\times3600s)^{-1}\times(1/6.52)\times10^{-6}\nicefrac{cm^{3}}{g}\times10m^{-1}=7.1\times10^{-6}s^{-1}$ } 
\begin{center}
\begin{table}
\begin{centering}
\begin{tabular}{|c|c|c|c|}
\hline 
Parameter & Value & Unit & Ref.\tabularnewline
\hline 
\hline 
$D_{0,w}$ & $1.6\times10^{-8}$ & $m^{2}/s$ & \cite{Holz_00}\tabularnewline
\hline 
$T_{s}$ & $215$ & $K$ & \cite{Holz_00}\tabularnewline
\hline 
$\gamma$ & $2.1$ & $[]$ & \cite{Holz_00}\tabularnewline
\hline 
$D_{o.ox}$ & $10^{-10}$ & $m^{2}/s$ & \cite{Youssef_14}\tabularnewline
\hline 
$Q_{ox}$ & $52$ & $kcal/mol$ & \cite{Youssef_14}\tabularnewline
\hline 
$S_{yc}$ & $[1.2-5.2]$ & $GPa$ & \href{https://www.azom.com/properties.aspx\%3FArticleID\%3D133}{azom}\tabularnewline
\hline 
$K$ & $[72-212]$ & $GPa$ & \href{https://www.azom.com/properties.aspx\%3FArticleID\%3D133}{azom}\tabularnewline
\hline 
$T_{0}$ & $600$ & $K$ & \cite{Reyes_18}\tabularnewline
\hline 
$T_{L}$ & $660$ & $K$ & \cite{Reyes_18}\tabularnewline
\hline 
$L$ & $50$ & $\mu m$ & \cite{Reyes_18}\tabularnewline
\hline 
$O_{0}$ & $62.5$ & $M$ & \cite{Bard_80}\tabularnewline
\hline 
$Z_{0}$ & $71$ & $M$ & \cite{Mejia_16}\tabularnewline
\hline 
$k$ & $7.1\times10^{-6}$ & $s^{-1}$ & \cite{Zhang_07}\tabularnewline
\hline 
\end{tabular}
\par\end{centering}
\caption{Simulation parameters.\label{tab:Parameters}}
\end{table}
\par\end{center}

Using the Equation \ref{eq:Diff} with the transition scale predicted
in Equation \ref{eq:sc} and the parameters given in Table \ref{tab:Parameters},
the real-time stress development in the oxide scale is computed in
Figure \ref{fig:SigmaPre} in versus the porosity values from partial
cracks. From the Equation \ref{eq:Diff} and the corresponding oxide
scale thickness Equation \ref{eq:s}, the post transition growth rates
has been illustrated in Figure \ref{fig:sPost} in higher porosities
values. The coupled growth regimes, are illustrated Figure \ref{fig:sPrePost}
with the corresponding sensitivity analysis for post transition regime.\footnote{Note the logarithmic scale in this graph. }
\begin{center}
\begin{figure}
\subfloat[Stress development. \label{fig:SigmaPre}]{\centering{}\includegraphics[height=0.17\textheight]{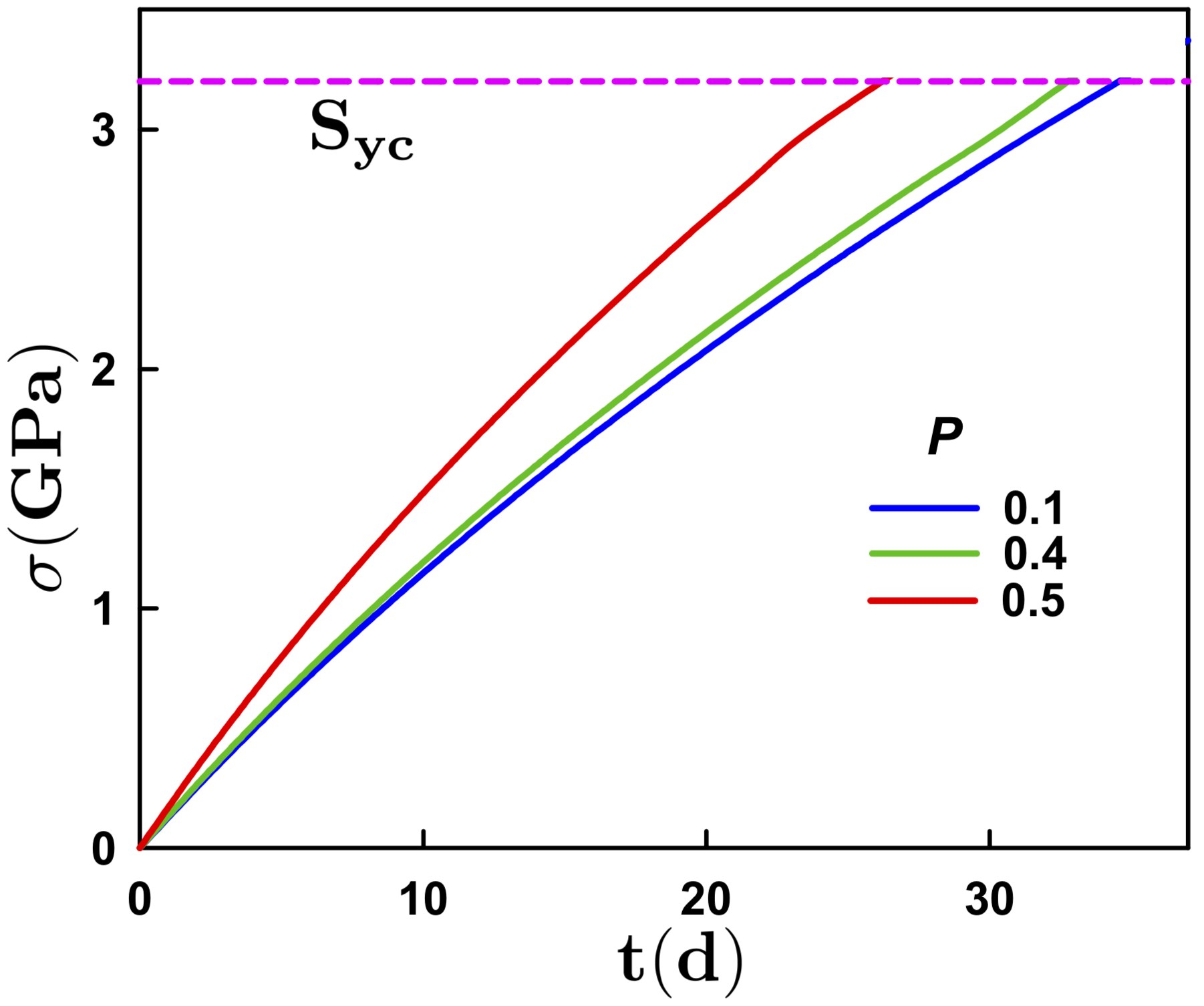}}\hfill{}\subfloat[Sensitivity analysis. ($b_{pre}=0.36$ ) \label{fig:sPrePost}]{\centering{}\includegraphics[height=0.17\textheight]{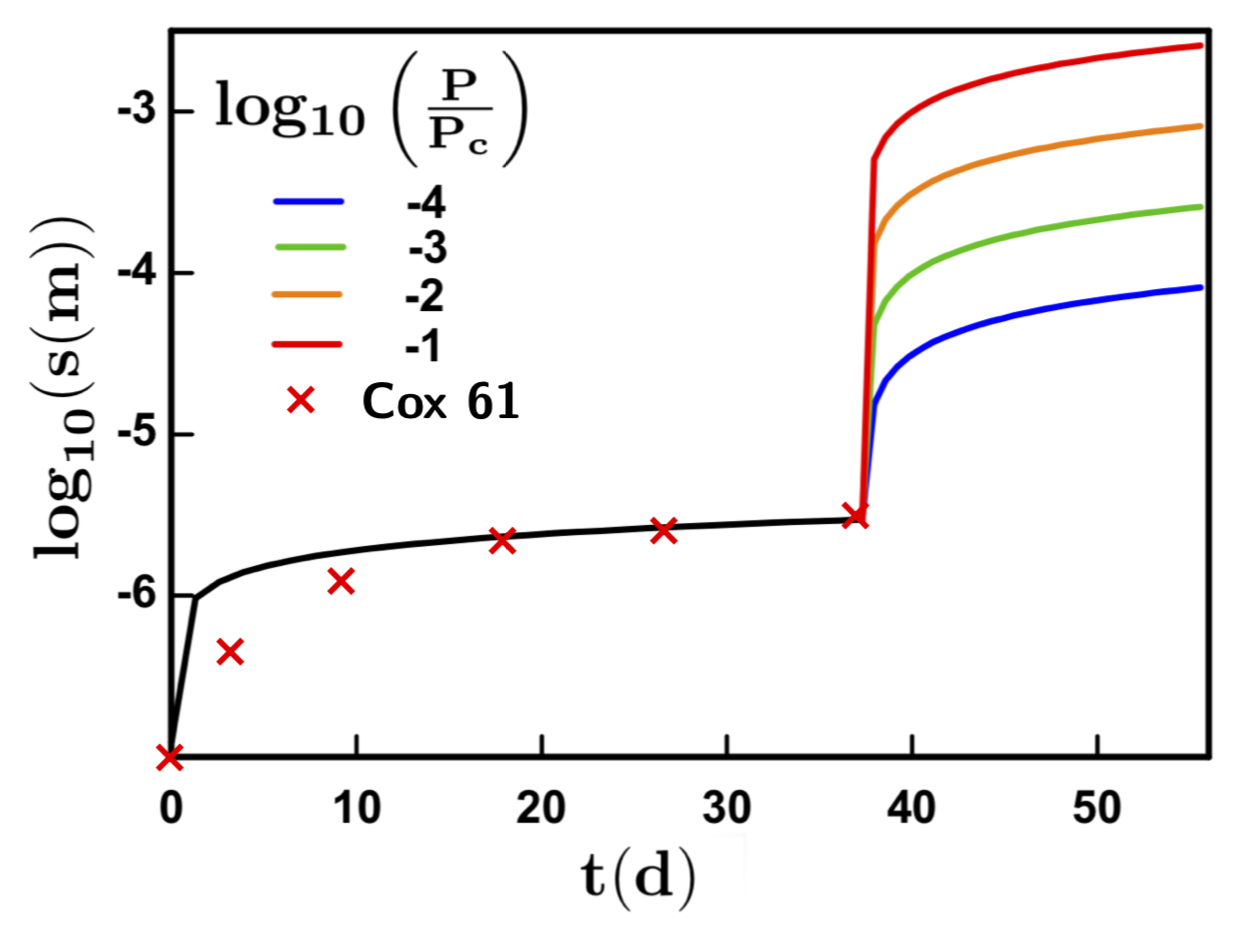}}\hfill{}\subfloat[Post-transition. \label{fig:sPost}($b_{post}\approx0.34)$]{\centering{}\includegraphics[height=0.17\textheight]{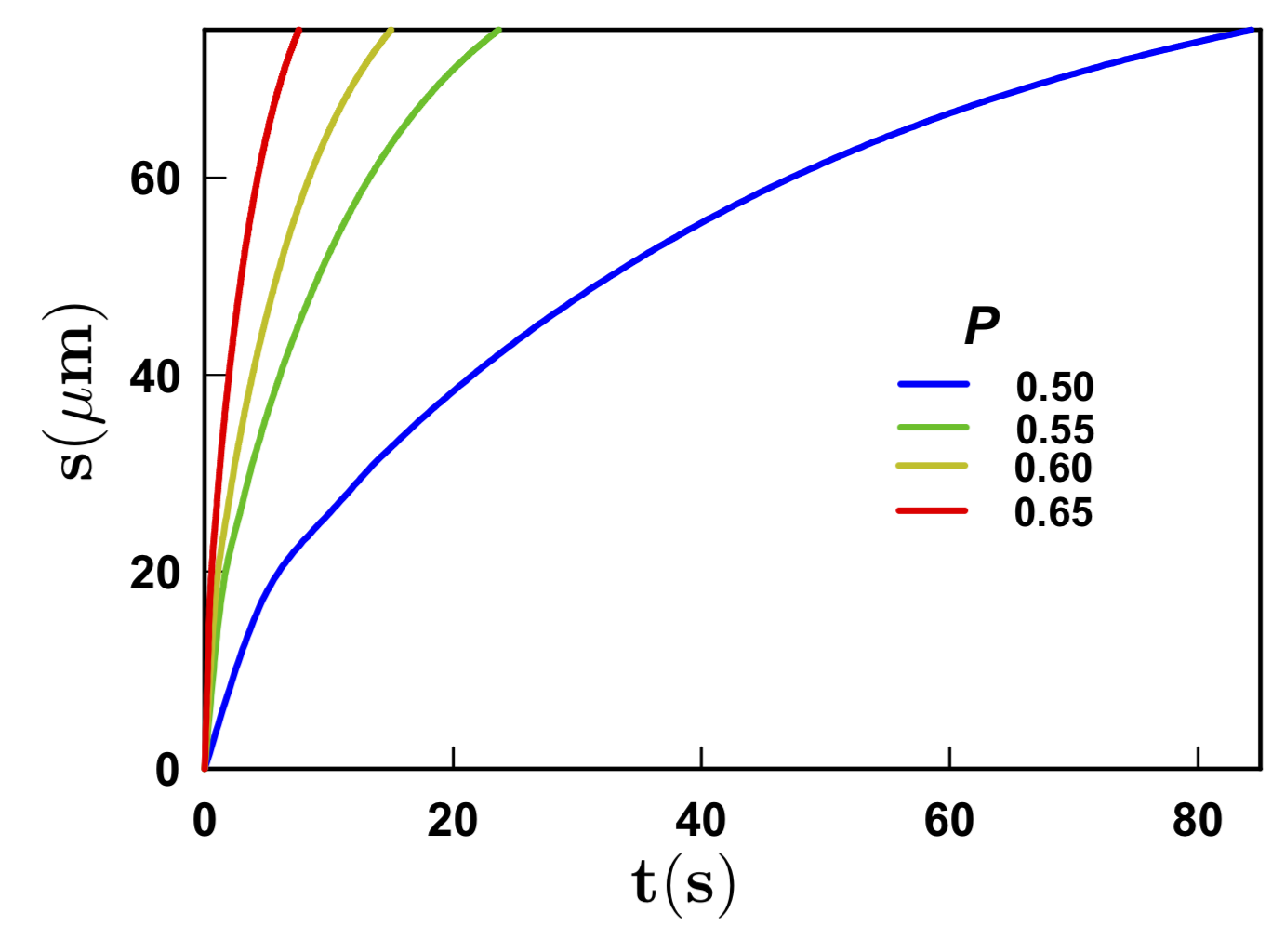}}

\caption{\textbf{(a)} Pre-transition growth regime. \textbf{(b)} Sensitivity
analysis growth regime during the post-transition regime. \textbf{(c)
}Growth regime after the fracture for difference porosity values.
\label{fig:s_t}}
\end{figure}
\par\end{center}

\subsubsection{Temperature profile:}

As the diffusivity values are very sensitive to the temperature, consideration
of it\textquoteright s distribution would be very useful realization.
Comparing the diffusivity of oxygen in zirconium ($D_{ox}\approx10^{-15}m^{2}/s$)
with thermal diffusivity of Zirconium ($\alpha\approx10^{-8}m^{2}/s$),
one ascribes: 

\[
\frac{\partial T}{\partial t}\gg\frac{\partial O}{\partial t}
\]
which implies that the kinetics of thermal propagation is significantly
higher and therefore at any infinitesimal period of time, temperature
profile has already reached the steady-state regime.(i.e. $T(y,t)\approx T(y)$
) As the oxide layer grows, it makes a relative insulation between
two ends, and consequently the heat flux ($q$) decreases, until fracture.
The enthalpy of the oxidation is negligible relatively to the amount
of transferred heat ($\Delta H_{rxn}\ll q$), therefore the heat flux
is controlled by the insulation of oxide layer: 

\begin{equation}
q(t)=-\kappa_{ox}\left(\frac{dT}{dy}\right)_{ox}=-\kappa_{Zr}\left(\frac{dT}{dy}\right)_{Zr}\leq-\kappa_{Zr}\frac{T_{L}-T_{0}}{L}\label{eq:q}
\end{equation}

The temperature at the oxide/metal interface ($T_{s}$) correlates
directly with their thermal conductivity and is inversely proportional
with the distance from each boundaries. Therefore if we define the
conductivity coefficient ($\beta:=\kappa/l$) . Therefore the interface
temperature ($T_{s}$) could be linearly obtained as:

\begin{equation}
T_{s}=\left(\frac{\beta_{ox}}{\beta_{ox}+\beta_{Zr}}\right)T_{0}+\left(\frac{\beta_{Zr}}{\beta_{ox}+\beta_{Zr}}\right)T_{L}\label{eq:T}
\end{equation}

where $\beta_{Zr}=\kappa_{Zr}/(L-s(t))$ and $\beta_{ox}=\kappa_{ox}/s(t)$
respectively. 

\subsubsection{Numerical stability:}

We utilize the finite difference scheme to solve the PDE Equation
\ref{eq:Diff} in space and time. If ($O_{i}^{j}$) represents the
oxygen concentration at depth ($y_{i}$) and time ($t^{j}$), adopting
forward difference method in time and space ($FTFS$), we get: 

\begin{equation}
O_{i}^{j+1}=(1-2Q_{1}-Q_{2}-Q_{3})O_{i}^{j}+(Q_{1}+Q_{2})O_{i+1}^{j}+Q_{1}O_{i-1}^{j}\label{eq:Oij}
\end{equation}

where $Q_{1},$$Q_{2}$ and $Q_{3}$ are the quotients defines as
below: 

\textbf{
\begin{equation}
Q_{1}=\frac{D\delta t}{\delta y}\label{eq:Q1}
\end{equation}
}

\begin{equation}
Q_{2}=\begin{cases}
\displaystyle{\Huge\frac{D\delta t}{\delta y}\frac{Q_{ox}}{RT^{2}}\frac{dT}{dy}} & \text{Pre}\\
\displaystyle{\Huge\frac{D_{0}\gamma\delta t}{\delta y}\left[\frac{T}{T_{0}}-1\right]^{\gamma-1}\frac{dT}{dy}} & \text{Post}
\end{cases}\label{eq:Q2}
\end{equation}

and 

\begin{equation}
Q_{3}=k\delta t\label{eq:Q3}
\end{equation}

where $\delta t$ and $\delta y$ are the segmentations in time and
space. To ensure the stability, we must have: 

\[
1-2Q_{1}-Q_{2}-Q_{3}\geq0
\]

Therefore, to satisfy the criteria for both growth regimes, the following
condition would suffice: 

\begin{equation}
\delta t:\leq\frac{\delta y^{2}}{2D}\label{eq:dtdy}
\end{equation}

\section{Results and Discussions}

The computational results demonstrate the cooperative role of diffusion
and migration for corrosion rate as well as significant sensitivity
of corrosion rate for the porosity values beyond percolation threshold.
Following the computational algorithm in flowchart \ref{fig:FlowChart}
the percolation pathways could be obtained through the crack network.
The density of states $M$ for partial and fully-connecting cracks
is shown in Figure \ref{fig:M}. It is obvious that upon increasing
the porosity $p$, the percolation through the partial cracks augments
earlier than the full cracks. Such a contrast also has been visualized
in the Figure \ref{fig:Sample}, where the spectrum of colors represent
the amount of time the oxygen has been present in those sites. Albeit
possessing less density of states, the full cracks have significantly
more impact on the oxide growth rate, merely due to much higher value
of water self-diffusion coefficient relative to the zirconium oxide. 

The method can predict the rate of corrosion from experimental images
by means of image processing. The bare image from cracked zirconium
oxide in Figure \ref{fig:Original} can be binarized via Otsu's minimization
of intraclass variance $\sigma^{2}$ in Equation \ref{eq:Otsu}. The
extracted $CRs$ are shown in Figure \ref{fig:Effective}. 

Performing scale studies, Figure \ref{fig:Scaling} shows that the
mean statistical density of states for the percolation samples, correlates
with the power law growth regime $1.9<\alpha<2.0$ as a verification.
\cite{Stauffer_94} Additionally in this graph, the bond percolation
curve stands higher than the site percolation. This comparison will
be verified mathematically in Equation \ref{eq:BondSiteComparison}.
Note that upon reaching the percolation threshold, the constriction
river most likely is the thinnest on the verge of percolation threshold
($p\approx p_{c}$). 

Given certain porosity value $p$, the water is transported through
a torturous pathway from the water/metal interface to reach the oxidation
front (i.e. oxide/metal interface). For lower values of porosity,
merely close to the percolation threshold $p\approx p_{c}$, there
will be a signifiant search for the shortest pathway due to scarcity
of the penetrable areas and the tortuosity value of the shortest path
will be the highest. As the porosity value increase, the possibility
of more direct connection is also becomes greater and therefore the
corresponding tortuosity value is reduced, such that in the limit
of full porosity ($p\rightarrow1$) the connection routes are almost
straight and the tortuosity value merges to unity. This trend has
been illustrated in Figure \ref{fig:TortPor} and shows a nice agreement
with previous findings. \cite{Koponon_96,Nabovati_07} Additionally,
in 2D percolation the transport is possible from 4 pathways, where
in 3D the percolation can occur from 6 directions, where, in each
case, only one direction is considered to be straight. Thus, the probability
of twisted percolation in 2D would be $\nicefrac{3}{4}$, whereas
in 3D it would become $\nicefrac{5}{6}$. This clearly shows that
the 3D percolation would generate more tortuosity in average versus
2D percolation, as shown in Figure \ref{fig:TortPor}.

The Diffusion coefficient values based on Equations \ref{eq:CrackSeries}
and \ref{eq:DDiscretized} are illustrated in Figure \ref{fig:DiffPor}.
Due to lower tortuosity values in 3D relative to 2D percolation (Figure
\ref{fig:TortPor}), it is obvious from Equation \ref{eq:CrackSeries}
that the diffusion coefficient also follows the same comparative trend.
Additionally, the values for bond percolation is more than site percolation,
hereby we prove that this is always true.

Given certain porosity $p$ , the bond percolation develops more diffusion
coefficient versus site percolation. The underlying reason is that
any square bond percolation paradigm ($4^{2}$), $p_{b}$ can be interpreted
by an equivalent bond percolation scheme ($4^{2}$), $p_{eq,s}$ where
each connection point (i.e. corner) could be treated as a site per
see. Assuming the dimension in a 2D bond percolation to be $d$ and
the number of available sites for percolation is $M$, the porosity
will be calculated as: 

\[
p_{b}=\frac{M}{d^{2}}
\]

where the dimension in the equivalent site percolation would be $2d+1$
, due to inclusion of connection points as available site.Therefore,
the elements in equivalent site percolation paradigm would be $n+(d+1)^{2}$,
and: 

\[
p_{eq,s}=\frac{M+(d+1)^{2}}{(2d+1)^{2}}
\]

Hence, we need to prove the following inequality: 

\[
\frac{M+d^{2}+2d+1}{4d^{2}+4d+1}<\frac{M}{d^{2}}
\]

performing rearrangements and since always $p_{b}<1$ , we have $M<d^{2}$
and we arrive at: 

\begin{align*}
d^{4}+2d^{3}-3Md^{2}-4Md-M+d^{2} & <\\
 & d^{4}+2d^{3}-3d^{4}-4d^{3}-d^{2}+d^{2}<0
\end{align*}

Therefore we only need to prove the RHS inequality, which gets simplified
into: 

\[
-2d^{2}(d+2)<0\text{ \ensuremath{\checkmark\ }}
\]

since this equation is always true, we have: 

\begin{equation}
p_{eq,s}>p_{b}\label{eq:BondSiteComparison}
\end{equation}

Thus, given a certain porosity, the bond percolation creates a larger
cluster (i.e. available sites) versus the site percolation, which
is obvious in Figure \ref{fig:DiffPor}.

For the transition regime based on compression stress, if the average
values are considered for the range of compressive yield strength
and th bulk modulus given in Table \ref{tab:Parameters}, the mean
value for the critical thickness for the transition state from Equation
\ref{eq:scValue} would be $s_{c}=2.75\mu m$ which is in nice agreement
with the values given in current literature. \cite{Causey_05,Cox_05}
Such transition state has been addressed with the close proximity
in recent findings with an alternative method as well. \cite{Reyes_18}

The large amount of Pillar-Bedworth ratio $R_{PB}$ indicates upon
formation and advancing oxide layer, the compressive stresses accumulate
in real time. We have captured such stress augmentation in Figure
\ref{fig:SigmaPre}, versus the pre-transition porosity $p$ (i.e.
partial cracks/ imperfections), until the yield limit (i.e. fracture).
It is obvious that the higher density of imperfections will reduce
the transition time. From Equation \ref{eq:Syc} the stress growth
$\sigma(t)$ depends to the oxide scale $s(t)$. This correlation
in particular is linear (i.e. direct) for higher values of original
thickness, where denominator will remain relatively invariant, and
therefore the cubic growth regime is expected (as will be explained
next) for stress growth behavior, where $s\ll L$. Nevertheless, the
real-time stress can be obtained from Equation \ref{eq:Syc} as: 

\[
\sigma(t)=2K\ln\left(\frac{R_{PB}-1}{R_{PB}}s(t)+1)\right)
\]

which shows exponential decay behavior as illustrated in Figure \ref{fig:SigmaPre}. 

Figure \ref{fig:sPrePost} illustrates the \emph{all-in-one} plot
for oxide evolution. The breakaway point has been analytically calculated
from Equation \ref{eq:scValue} and the sensitivity analysis has been
performed for the post-transition regime based on the logarithmic
distance from percolation threshold. The growth regime in either stage
can be approximated with the power law growth in time as: 

\begin{equation}
s(t):=at^{b}\label{eq:PowerLaw}
\end{equation}

Correlating the pre-transition growth regime with power law, the exponent
value of $b_{Pre}=0.35$ is obtained, which is in very high agreement
with the value of $b\approx\nicefrac{1}{3}$ in the literature. \cite{Motta_09,Cox_60,Reyes_18,Allen_12}
On order to verify the simulation results, we use corresponding experimental
results from \cite{Cox_60}, where the the increase in the mass of
oxide samples have been correlated with a cube of time and within
38 days of corrosion has possessed the weight addition of $18\nicefrac{mg}{dm^{2}}$
\footnote{Ref. \cite{Cox_60}, Figure 2.} which translates to the
oxide thickness value of $2.76\mu m$.\footnote{18$\cfrac{mg}{dm^{2}}=18\times\cfrac{10^{-3}}{10^{2}}\cfrac{g}{cm^{2}}=1.8\times\cfrac{1}{6.52}\times10^{-5}m=2.76\mu m$}

For fundamental understanding of cooperating role of diffusion, reaction
and thermomigration terms, one can start from the typical diffusion
Equation given as below: 

\begin{equation}
\frac{\partial O}{\partial t}=\frac{\partial}{\partial y}\left(D\left(\frac{\partial O}{\partial y}\right)\right)\label{eq:NormalDiffusion}
\end{equation}

The typical solution for this Equation will have a parabolic trend.
($s\propto t^{\nicefrac{1}{2}}$). The subtraction of significant
consumption term ($-kO$) with the presence of thermomigration term
$D\alpha(T)\cfrac{\partial O}{\partial y}$ in Equation \ref{eq:Diff},
would bend down (i.e. reduce) the oxide evolution curve, such that
in our simulations it correlates with a lower power value. (i.e. $b\approx0.36$).
The aforementioned bending effect has been addressed by presence of
exponentially reducing electric charge distribution and the corresponding
electrostatic field at reaction sites during previous study.\cite{Reyes_18}

Additionally, Figure \ref{fig:sPrePost} addresses the extremely high
sensitivity of the growth kinetics upon reaching the percolation threshold.
($p_{bond}=0.5$ , $p_{site}=0.5928$) as plotted in logarithmic scale.
\cite{Vidales_95} In fact, during very the initial moments of post-transition
regime, since there is abundance of oxygen in the reaction sites,
the oxide growth will be merely \emph{reaction-limited. }Therefore
the transport regime will be negligible and the growth in the oxide/metal
interface can be approximated by: 

\begin{equation}
\left(\frac{\partial O}{\partial t}\right)_{post,0}\approx-kO\label{eq:DiffPost}
\end{equation}

Note that due in the interface of oxide and metal with the infinitesimal
thickness $\delta y$ , and any other cracked region, the evolving
concentration of oxygen $O$ initially is only a function of time.
Equation \ref{eq:DiffPost} can be solved analytically as below: 

\[
O_{post,0}=-c_{1}(y)\exp(-kt)+c_{2}(y)
\]

During the initial moments, the oxygen is already available in the
reaction sites through cracks, therefore:

\[
O(0)=O_{0}
\]
 On the other hand, the Equation should should satisfy itself in the
initial moment. Hence, from the two boundary conditions, we have:
$c_{1}=O_{0}$ and $c_{2}=0$ and the time-dependent initial concentration
profile turns to be: 

\[
O_{post,0}=O_{0}\exp(-kt)
\]

In order to obtain the initial oxide thickness $s_{post,0}(y,t)$
we can integrate this concentration profile based on Equation \ref{eq:s} where
the developed oxide thickness turns to be: 

\begin{equation}
s_{post,0}(y,t)=\frac{O_{0}y}{2Z_{0}}\exp(-kt)\label{eq:sReactive}
\end{equation}

Such initial exponential decay regime in time has been also addressed
in the past.\cite{Korobkov_58} Note that the initial profile is linear
versus depth $y$ which is also shown in the literature.\cite{Cox_60}
Such a initially linear growth regime could be also discerned during
both growth regimes in Figure \ref{fig:sPrePost}. Nevertheless, our
understanding from post-transition growth regime is via developing
our analytical methods due to lack of research in the literature.

As the oxide layer evolves the diffusive term $D(\frac{\partial^{2}O}{\partial y^{2}}+\alpha(T)\frac{\partial O}{\partial y}$)
becomes relatively more significant due to depletion and scarcity
of oxygen in reaction sites, where the growth regime turns to be \emph{diffusion-limited}.
Such a growth regime correlated with square root of time ($s\sim t^{\nicefrac{1}{2}}$).
In fact the coupled diffusion-reaction (i.e. consumption) evolution
of oxide scale is approximated with a power-law growth curve in time
($c_{2}(y)t^{b}$) (Equation \ref{eq:PowerLaw}), which is below the
growth by sole diffusion and above the the growth by sole reaction,
therefore one expects the following:

\begin{equation}
c_{1}(y)\exp(-kt)<c_{2}(y)t^{b}<c_{3}(y)t^{\nicefrac{1}{2}}\label{eq:Ineq}
\end{equation}

where $\{c_{1}$, $c_{2}$ ,$c_{3}\}>0$ are time-independent coefficients.
Given large-enough time, the role of the coefficients in the inequality
becomes negligible and in order for the Equation \ref{eq:Ineq} to
be always true, we must have: 

\[
0<b<\nicefrac{1}{2}
\]

Which is addressed throughout the literature and during this study.
\cite{Adamson_07,Beie_94,Belle_54} In fact the power coefficient
$b$ should express the cooperative interplay between the \emph{corrosion-assisting}
diffusion and \emph{corrosion-resisting} reaction terms. 

Anther important factor for the diffusion-reaction development is
to ensure that there is always oxygen available for consumption in
the reaction sites, before reaching the stoichiometric (i.e. saturation)
limit. In other words, the transport (i.e. diffusive) term of oxygen
should always be competitive with the reactive (i.e. consumption)
term. Such juxtaposition during large time intervals can be qualitatively
expressed as below: \footnote{By means of Taylor expansion, the exponential term can be expressed
as: $\exp(-kt)=1-kt\cancelto{0}{+O(k^{2})}$, where the second order
term $O(k^{2})$ is negligible due to very small value of reaction
constant $k$. (Table \ref{tab:Parameters}) }

\[
\sqrt{2}D^{\nicefrac{1}{2}}t^{\nicefrac{1}{2}}>\cfrac{O_{0}y}{2Z_{0}}\exp(-kt)\approx\cfrac{O_{0}y}{2Z_{0}}(1-kt)
\]

where $\sqrt{2}D^{\nicefrac{1}{2}}t^{\nicefrac{1}{2}}$ is the mean
square displacement of the diffusion interface \cite{Aryanfar_18}
and the \emph{RHS} is the movement of the reactive interface, given
in Equation \ref{eq:sReactive}. Re-arranging this equation yields
to the following dimension-free inequality: 

\begin{equation}
kt+qt^{\nicefrac{1}{2}}-1>0\label{eq:Gamma}
\end{equation}

where $q:=\cfrac{2Z_{0}\sqrt{2D}}{O_{0}y}$ is the coefficient\footnote{with the unit of $s^{\nicefrac{-1}{2}}$.}.
The Equation \ref{eq:Gamma} is quadratic has a real root ($\Delta=q^{2}+4k>0$)
given below:

\begin{equation}
t_{reac\rightarrow diff}=\left(\frac{-q+\sqrt{q^{2}+4k}}{2k}\right)^{2}\label{eq:TransitionDiffReac}
\end{equation}

which means that there is a critical time-interval $t_{reac\rightarrow diff}$
before which the consumption is dominant (i.e. reactive) and after
that the concentration accumulation occurs. Such transition from \emph{reaction-limited}
to \emph{transport-limited} depends on variables forming $q$. In
other words the transition takes the longest, $q$ is smaller or comparable
with $k$. This is very reasonable since the higher depth values usually
\emph{'suffer breathing'} due to lack of oxygen inflow. Additionally,
the higher values of reaction constant $k$ will cause more consumption
rate and will augment this value respectively. On the interface, where
$y\rightarrow0$, we have $q\gg k$ and the transition is immediate
(i.e. $t_{reac\rightarrow diff}\approx0$)

Figure \ref{fig:sPost} represents the growth regime of the oxide
scale for the porosity values beyond the fracture porosity ($p>p_{c}$).
The most distinct curve here is, in fact, in the vicinity of percolation
threshold ($p\approx p_{c}=0.5$ ). Assuming to maintain the same
porosity upon fracture, the power coefficient has been obtained as
($b_{post}\approx0.34$). The decrease in the power value relative
to can be interpreted as the negligence of the thermomigration during
the post-transition period, which in fact will help to increase the
term $\frac{\partial O}{\partial t}$ relative to pre-transition regime.
In fact, the post-transition growth regime, can be interpreted a second
pre-transition regime, where the oxygen has made ways through the
reaction site and therefore the growth regime is as expected. Additionally,
this is also the underlying reason for \emph{quasi-cycling} growth
behavior (i.e. multiple oxidation and fracture stages of zirconium
and its alloys) throughout corrosion, which has been addressed in
numerous places in the literature. \cite{Hillner_2000,Allen_12}

\section{Conclusions}

In this paper, we have developed a constriction percolation paradigm
for the pre- and post-transition growth regime of zirconium, distinguishing
the transition by means of when the crack density $p$ meets the percolation
threshold $p_{c}$. Consequently we have established a coupled diffusion-reaction
framework to predict the growth regime throughout the corrosion event,
extending beyond fracture. We have verified the results by means of
literature, the contrast between the square site and bond percolation
methods and analytical methods. Additionally we have developed a formulation
for compression-based yielding of zirconium to predict the onset of
the transition. In particular, we have proved that there is a critical
time, in which the corrosion event moves from reaction-limited oxide
evolution to diffusion-limited growth regime and we have analytically
described the range of the power coefficient for the power-law growth
kinetics. 

\subsection*{Acknowledgement}

This research was supported by the Consortium for Advanced Simulation
of Light Water Reactors (CASL), an Energy Innovation Hub for Modeling
and Simulation of Nuclear Reactors under U.S. Department of Energy
Contract No. DE-AC05-00OR22725. The authors also acknowledge providing
the sample experimental image from Dr. Joe Rachid at Pacific Northwest
National Lab (PNNL). 

\bibliographystyle{unsrt}
\bibliography{Refs}

\end{document}